\documentclass[conference,compsoc]{IEEEtran}
\usepackage[nocompress]{cite}
\usepackage{amsmath,amssymb,amsfonts}
\usepackage{algorithmic}
\usepackage{graphicx}
\usepackage{booktabs}
\usepackage{caption}
\usepackage{subcaption}
\usepackage{textcomp}
\usepackage{xcolor}
\usepackage{float}
\usepackage{array}
\usepackage{url}
\usepackage[symbol]{footmisc}
\usepackage[small,compact]{titlesec}

\DeclareRobustCommand*{\IEEEauthorrefmark}[1]{%
  \raisebox{0pt}[0pt][0pt]{\textsuperscript{\footnotesize #1}}}

\iftrue
\def\TU#1{}
\def\RV#1{}
\def\NF#1{}
\def\HL#1{}
\def\JK#1{}
\def\MK#1{}
\fi

\begin{document}
 
\title{Toward an Automated HPC Pipeline for Processing Large Scale Electron Microscopy Data }
\iftrue
\author{\IEEEauthorblockN{Rafael Vescovi\IEEEauthorrefmark{1}\textsuperscript{*}, 
Hanyu Li\IEEEauthorrefmark{2}\textsuperscript{*}, 
Jeffery Kinnison\IEEEauthorrefmark{3}, 
Murat Ke\c{c}eli\IEEEauthorrefmark{4}, \\
Misha Salim\IEEEauthorrefmark{1},
Narayanan Kasthuri\IEEEauthorrefmark{2},
Thomas D. Uram\IEEEauthorrefmark{1} and
Nicola Ferrier\IEEEauthorrefmark{5}}
\IEEEauthorblockA{\IEEEauthorrefmark{1}Leadership Computing Facility, Argonne National Laboratory, Lemont, IL, USA\\
ravescovi,msalim,turam@anl.gov}
\IEEEauthorblockA{\IEEEauthorrefmark{2}Department of Neurobiology, University of Chicago, Chicago, IL, USA\\
hanyuli,bobbykasthuri@uchicago.edu}
\IEEEauthorblockA{\IEEEauthorrefmark{3}Dept. of Computer Science and Engineering, University of Notre Dame, Notre Dame, IN, USA\\
jkinniso@nd.edu}
\IEEEauthorblockA{\IEEEauthorrefmark{4}Computational Science Division, Argonne National Laboratory, Lemont, IL, USA\\
keceli@anl.gov}
\IEEEauthorblockA{\IEEEauthorrefmark{5}Mathematics and Computer Science Division, Argonne National Laboratory, Lemont, IL, USA\\
nferrier@anl.gov}}

\fi

\maketitle

\renewcommand{\thefootnote}{\fnsymbol{footnote}}
\color{white}
\footnote[1]{These authors contributed equally to this work.}
\color{black}
\begin{abstract}
We present a fully modular and scalable software pipeline for processing electron microscope (EM) images of brain slices into 3D visualization of individual neurons and demonstrate an end-to-end segmentation of a large EM volume using a supercomputer. Our pipeline scales multiple packages used by the EM community with minimal changes to the original source codes. We tested each step of the pipeline individually, on a workstation, a cluster, and a supercomputer. Furthermore, we can compose workflows from these operations using a Balsam database that can be triggered during the data acquisition or with the use of different front ends and control the granularity of the pipeline execution. We describe the implementation of our pipeline and modifications required to integrate and scale up existing codes. The modular nature of our environment enables diverse research groups to contribute to the pipeline without disrupting the workflow, \textit{i.e.} new individual codes can be easily integrated for each step on the pipeline. 
\end{abstract}


\section{Introduction}

Microscopy images are a significant source of insight and raw information for neuroscience. Modern techniques in electron microscopy (EM) allow scientists the ability to image at such high resolution that every single synaptic connection can be distinguished \cite{briggman2006towards, helmstaedter20083d, kasthuri2015saturated}. Furthermore, acquisition automation has enabled us to acquire large volumes of microscopy data spanning several resolutions with minimal human involvement in the acquisition.

The popularization of these automated imaging systems has made acquiring large amounts of data the standard operation for many laboratories \cite{rubin2006janelia}~\cite{sunkin2012allen}, and although most of them are able to physically handle the amount of data, there is an increasing need for streamlining the pipeline. This necessity arises because of the growing acquisition speed of microscopes, leading to an exponential growth in data throughput\cite{keller2014high}. While different techniques are emerging to solve each step of the upstream process, they still have their own independent development communities\cite{o2018large,cardona2012trakem2, wetzelhigh} and there are very few laboratories with the capability of carrying out the entire process by themselves. Individually these  algorithms contribute to the study of neuroscience image data; it is, however, non-trivial to chain these modules together and deploy them in one coherent environment for end-to-end connectomics projects. 

The continuous use of electron microscopes can produce single datasets that reach multiple petabytes of data, which cannot be processed on local workstations or small clusters and therefore require High Performance Computing (HPC) facilities. We propose to deploy and chain these different processing libraries into a single \textit{microscope to HPC} workflow and provide a way for the user to interact \HL{monitor? it's not clear what interact means here} with the data and its processing in real time. Our package is implemented in Python and is called HAPPYNeurons (HPC Automated Pipeline for Processing Yotta Neurons) \cite{repo}. 

An alternative solution is the use of cloud-based services (e.g. AWS, Google Cloud etc.), but the data size and computation time make the cost infeasible for most laboratories. On the other hand, HPC facilities at national labs have vast storage and computing capabilities that have not been tapped in the field of connectomics due to accessibility or differences from established technology stacks. Our pipeline aims to address this limitation by incorporating state-of-the-art open source connectomics tools into an HPC environment and building an end-to-end pipeline for EM segmentation, while increasing accessibility to users and labs.

On a technical level, we propose to use MPI \cite{gropp1999using} as a parallelization layer for each step in order to keep the internal mechanics of the original software mostly intact, while achieving compatibility with most HPC infrastructure. This allows for rapid deployment of new tools on the pipeline.
It is not the goal of our work to improve the sample preparation or acquisition, but instead to enable the user to take advantage of large scale computing facilities in order to process the data. This processing is currently done in several parts that will be described in greater detail in the next chapter. 
Our contributions to the field of neuroscience and HPC are: 1) deployment on HPC of EM tools necessary to go from raw images to final reconstruction; 2) wrapping the tools in an operation database that can be used to create custom pipelines, 3) deployment of an computational environment that permits the user to interact with, annotate, and visualize the data without the need to transfer outside of the HPC facility.

\section{Related work}  

The field of \textit{connectomics} has been blooming with different communities trying to understand the underlying connectivity map of neural tissue. In the particular case of electron microscopy there have been parallel contributions into sample preparation, data acquisition and the diverse steps on the complex data processing involved. The sheer complexity of the problem means every step of it can still be improved and ongoing efforts on different aspects of the problem can be seen from various labs and groups in the community \cite{schneider2020chandelier, motta2019dense, kornfeld2020anatomical}. 
 
For alignment, computer vision methods
\cite{o2018large,cardona2012trakem2, wetzelhigh} are still heavily used to assemble the raw data into 3D volumes, but are slowly being surpassed by machine learning-driven methods \cite{mitchell2019siamese}. Segmentation has long been the rate-limiting step and still requires a lot of human annotation. Originally it could take weeks to trace a single neuron through a stack of images. With the aid of deep learning algorithms, segmentation can be vastly accelerated.  Efforts on neural networks such as U-Net~\cite{wu2019chunkflow} and Flood-Filling Network (FFN)~\cite{januszewski2018high} have proven successful for the task of automatic segmentation of neurons.

Human-intensive data annotation is crucial to establish datasets for machine learning approaches. The web-based package webKnossos \cite{boergens2017webknossos} enables laboratories to deploy an intuitive interface to annotate datasets, without requiring annotators to transfer up to petascale datasets between their institution and the hosting site. We leverage webKnossos in our work to make the increasingly large connectomics datasets available for annotation by distant annotators.

Upon completion of reconstruction, datasets are meshed for  visualization and made available via  Neuroglancer\cite{neuroglancer}, a program developed by Google that visualizes flat, black-and-white electron images, related labels, and reconstructions as a colourful 3D forest of neurons.

\section{Computational Pipeline}

\begin{figure}[ht!]
    \centering
    \includegraphics[width=\linewidth]{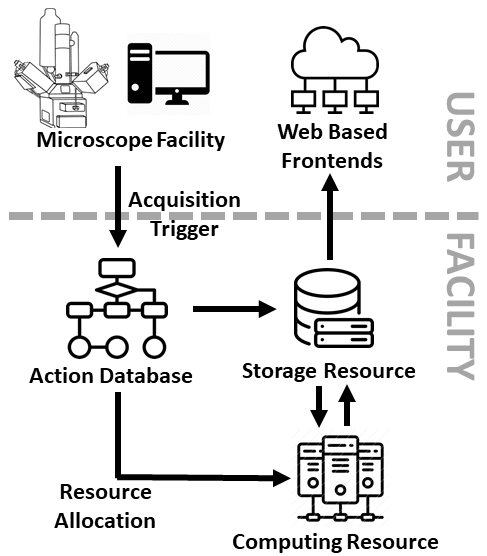}
    \caption{Model of the flow of data between the electron microscope facility (top side) and the HPC facility (lower side). The microscope acquisition populates the action database which controls the storage and computing resources. The user can visualize and manipulate the data and actions through web-based front ends.}
    \label{fig:steps}
\end{figure}

Electron microscope image processing follows a certain number of pre-defined steps from raw data to a final scientific result. Given the data throughput of modern microscopes, executing all those steps within the same facility as the microscope is a challenge. Figure \ref{fig:steps} describes the connection between the electron microscope lab at Argonne National Laboratory and the Argonne Leadership Computing Facility (ALCF), showing the services involved in our pipeline environment. After the microscope finishes acquiring an image (or set of images), it triggers an action that is stored on an external database server. This database of actions then controls the transfer and processing of data through the storage and computing resources. On the front end side the user can either manipulate the data or the actions to make the pipeline unique to each sample.

For simplicity we will describe the basic steps involved in processing electron microscope images individually. Even though these steps are described in a sequential fashion, in practice the processing is conducted through an iterative process involving automated computational steps and human-intensive guidance. 

\subsection{Processing methods for EM data}

In order to make the the interaction with the data as easy as possible for the user, we encoded basic operations that can eventually be described as sequential pipelines. Since human validation is necessary at multiple steps, the user can choose where and when to interact with the pipeline. These operations are described below:

\noindent\textbf{Montage} is the process of positioning and merging overlapping image tiles into a single larger image. Given sufficient metadata about the arrangement of the tiles, this step can be executed largely without user intervention. We implemented a headless macro for TrakEM2 \cite{cardona2012trakem2} and developed a Python wrapper for MPI parallelization of this procedure. The mutual independence of image sections makes it possible to trigger the montage operation during acquisition, once the full set of images for a tile have been acquired, and process it on the fly.  
    
\begin{figure}[h!]
    \centering
    \begin{subfigure}[t]{0.32\columnwidth }
        \centering
        \includegraphics[width=\linewidth]{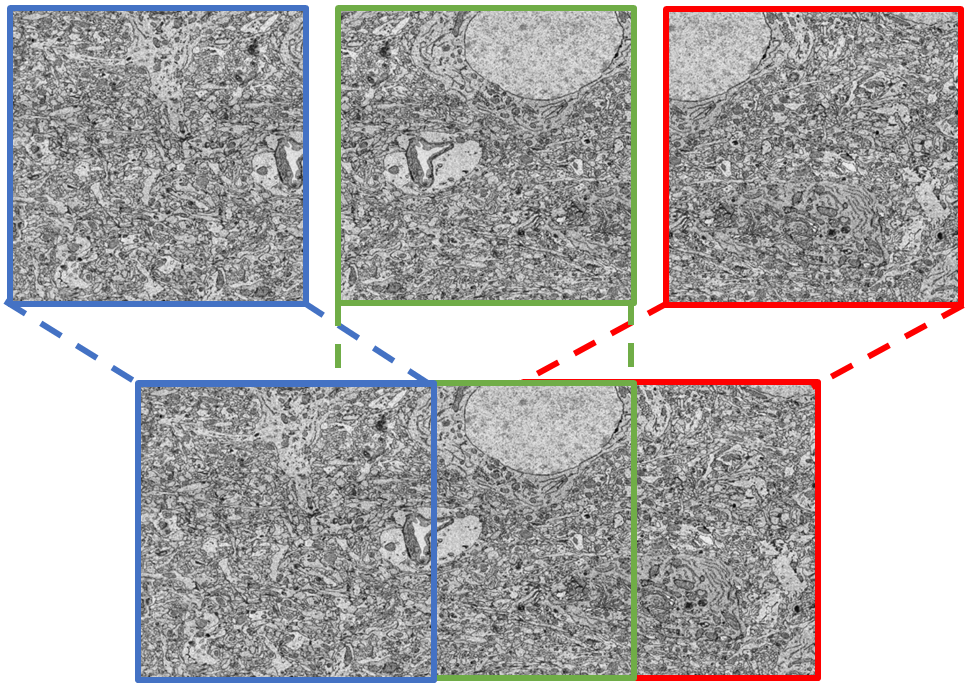} 
        \caption{Montage \\ Example} \label{fig:mont}
    \end{subfigure}
    \hfill
    \begin{subfigure}[t]{0.66\columnwidth }
        \centering
        \includegraphics[width=\linewidth]{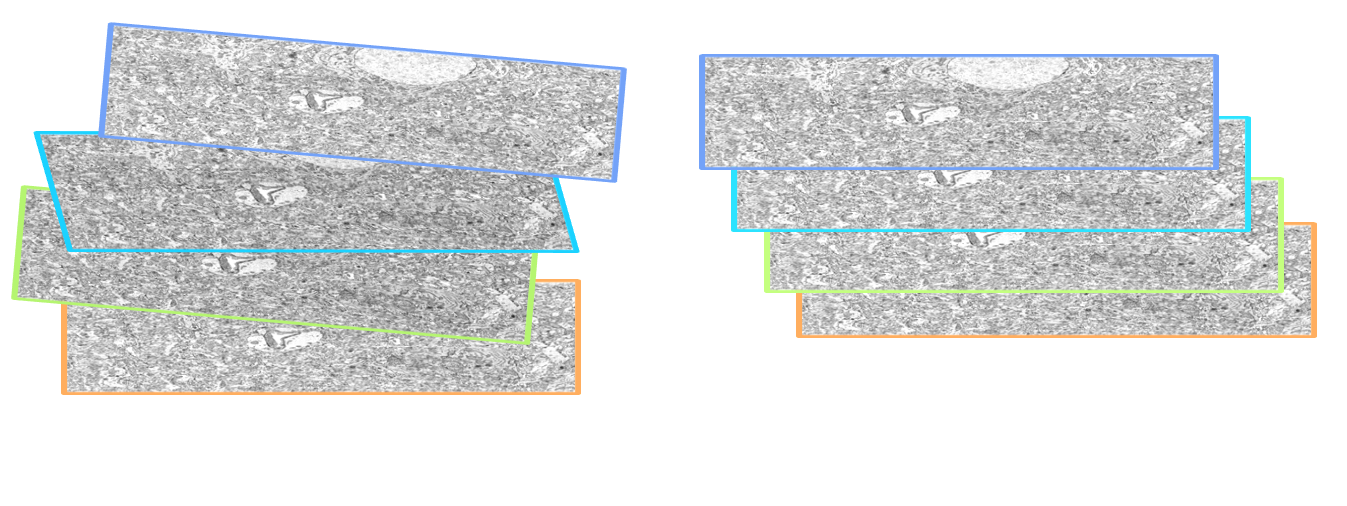} 
        \caption{Alignment example} \label{fig:align}
    \end{subfigure}
        \caption{Example of the montage process (left) and alignment process (right).}
\label{fig:mont_al}
\end{figure}

\noindent\textbf{Alignment} is the process of ensuring that neighboring images in the stack are aligned according to their contained features,  and is a crucial step for serial electron microscopy. In our pipeline, we use AlignTK\cite{wetzelhigh} to perform elastic alignment on the  montaged image stack. We implemented wrappers for AlignTK's core functionality to better adapt to parallel deployment on HPC, along with a set of utility tools for image preprocessing, including contrast normalization, scaling, and artifact thresholding. Figure \ref{fig:mont_al} shows examples of montage and alignment images.

\noindent\textbf{Segmentation} is the process to assign unique IDs to individual neurite objects in a 3D volume and is one of the key challenges in connectomics analysis. The current state of the art, FFN, has achieved great success in accuracy and scale \cite{januszewski2018high, zheng2018complete, kornfeld2020anatomical} over the last few years. Although it was originally designed for distributed computing platforms and despite preliminary efforts on distributed training\cite{dong2019scaling}, it has not previously been deployed on HPC infrastructure for large scale segmentation. In this pipeline, we made modifications to the open source release of Google's FFN. First, we added MPI-based parallelization for execution at large-scale HPC facilities. Second, we added support for reading precomputed volume\cite{silversmith2018cloudvolume} data as input, in addition to \textit{HDF5} which reduces repetitive data usage and seamlessly integrates with the visualization engine Neuroglancer. Third, we implemented a 
reconciliation step that merges overlapping subvolume inference results into a final segmentation in precomputed format.

\begin{figure}[h!]
    \centering
    \begin{subfigure}[t]{0.49\columnwidth }
        \centering
        \includegraphics[width=\linewidth]{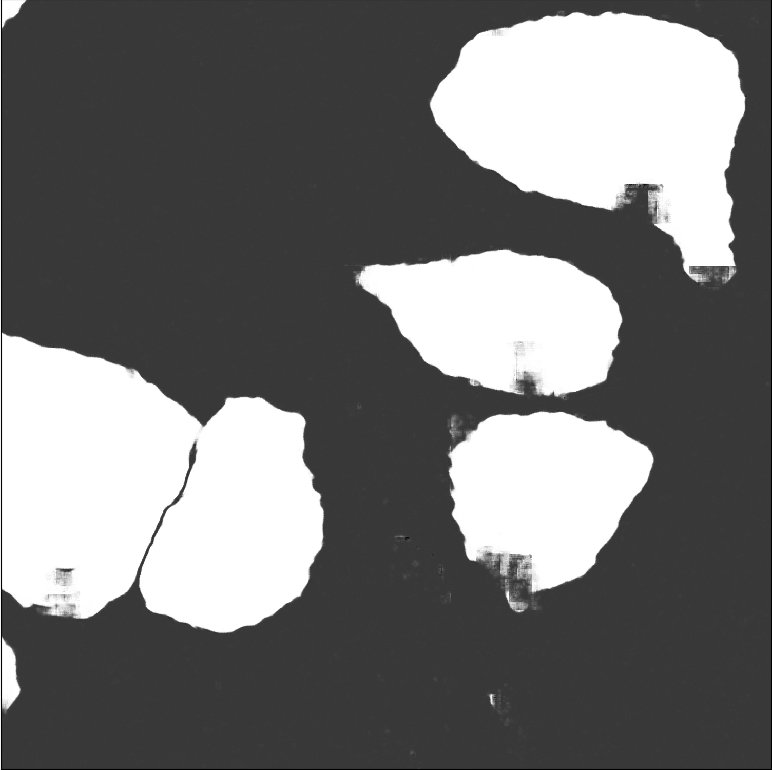} 
        \caption{U-Net cell body probabilities} \label{fig:unet1}
    \end{subfigure}
    \hfill
    \begin{subfigure}[t]{0.49\columnwidth }
        \centering
        \includegraphics[width=\linewidth]{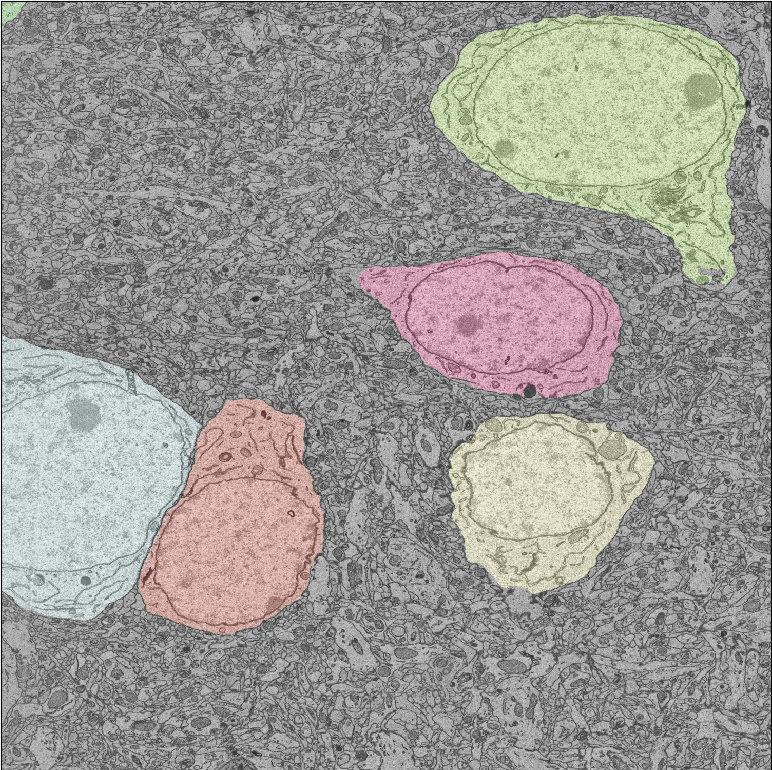} 
        \caption{U-Net Watershed Overlay} \label{fig:unet2}
    \end{subfigure}
    \caption{Result of the U-Net segmentation of large body sizes. These can be subsequently used to mask-out already known objects in the final segmentation.}
    \label{fig:unet_full}
\end{figure}

\noindent\textbf{Mask Prediction:} In practice, a prerequisite for FFN is identifying tissue masks that would disrupt segmentation. These masks are used to omit imaging artifacts or large objects that can make the final segmentation less accurate, like cell-bodies and blood vessels. For cell-bodies and blood vessels, we implemented and ran a classic 2D U-Net\cite{ronneberger2015u}. We created manual annotations on every 100 images at 4x resolution and used these to train a U-net model, which was used for patch-wise inference over the full volume. Manual seeds were placed at the center of each cell-body, and a 3D watershed algorithm was run to provide initial segmentation of cell bodies and blood vessels. Figure \ref{fig:unet_full} shows the process of creating the cell mask from the U-Net probabilities (on the left) to the actual cell body masks (on the right).

\noindent\textbf{Mesh Generation} produces a mesh-based representation to support 3D visualization of the segmented objects. Currently this step is achieved using the Python library Igneous\cite{igneous}.
 
\noindent\textbf{Skeletonization} creates a point graph for every object and can also be processed by using the TEASAR\cite{sato2000teasar} implementation inside Igneous.

\noindent\textbf{Manual Annotation:} We used WebKnossos\cite{boergens2017webknossos} to provide manual volumetric annotations as a training/validation set for FFN. This step is human-intensive, and is typically approached iteratively, with the biologist annotating an initial sample partially, rerunning training and inference with FFN, and making further corrections. To accommodate data format requirements of different packages and to streamline this process, we implemented utilities for easy transformation of data formats between \textit{WebKnossos cube}, which is used by WebKnossos, stacks of tiff images, and \textit{HDF5}\cite{folk2011overview}, which are traditional data formats, and \textit{precomputed}\cite{silversmith2018cloudvolume}, which is used by CloudVolume and Neuroglancer. 

User inspection of intermediate results between pipeline stages is currently essential. To this end, we have developed a number of facilities to enable interaction with the datasets that reside on ALCF systems. In the case of connectomics datasets, the intermediate results can be very large, typically involving many images in the range of hundreds of megapixels each; the ability to view these results quickly, in-place on ALCF systems, is critical. For each intermediate dataset, we have developed code to produce downsampled versions of select output data, and a Jupyter notebook template which can be copied into the target run directory to view the downsampled data. The JupyterHub deployment at ALCF has direct access to the Theta and Cooley filesystems, creating a highly usable environment for viewing, which can be customized with additional Python-driven analyses.

While segmentation results can be viewed using Jupyter notebooks as above, these results are more typically visualized in 3D using the Neuroglancer application. We have deployed Neuroglancer at ALCF to support viewing segmentation results, which we demonstrate later in the text. Whereas Jupyter notebooks access Theta-resident data directly, Neuroglancer retrieves data using web protocols; to support this, we transfer segmentation results to the 3PB Petrel community storage system\cite{allcock2019petrel} at ALCF, and expose the data to Neuroglancer using standard web protocols. 

\subsection{Workflow Management}

\begin{figure*}[ht!]
    \centering
    \includegraphics[width=14.5cm
    ,keepaspectratio,]{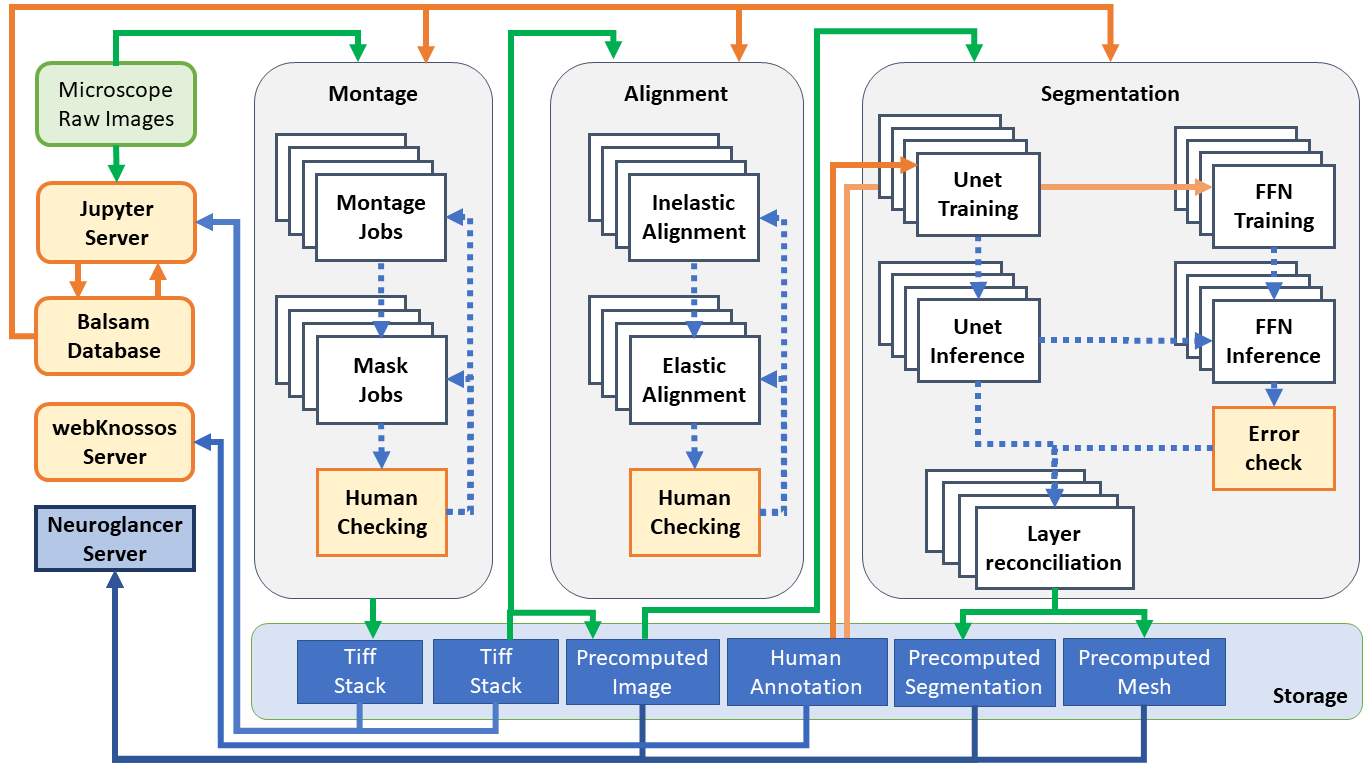}
    \caption{Schematic of the pipeline. The green box represents the data acquisition. Orange boxes and arrows represent human interactions in the pipeline. The white boxes represent HPC submitted jobs. Blue boxes represent the data storage and visualization server. Green arrows represent I/O from the computing resources.}
    \label{fig:comp_pipes}
\end{figure*}

We have designed a pipeline that encompasses the individual software packages described above, allowing them to be executed independently or assembled into more automated workflows. These software packages are leveraged in an iterative fashion, varying parameters to achieve desired accuracy on each particular dataset, and software development is ongoing. To account for this scenario, our pipeline is modular, supports multiple interfaces, and aims to enable the entire data life cycle from raw images to final results.  Figure \ref{fig:comp_pipes} shows the electron microscope operations described above with the I/O present on our package. 

Given the complex nature of the acquired data, we acknowledge that human intervention is required during the process; our goal is to facilitate this while  minimizing the iteration time and making interaction with the data easier. This is important given that any imperfection or artifact in the sample and data acquisition can cause the steps of the pipeline to fail and require human intervention. 

The pipeline is implemented as a Python library, with an API that exposes the individual applications to be run, together with specifications of input data and configuration, producing a collection of jobs that will be offloaded to our HPC facilities. This functionality is generalized such that choice of execution machine can be made at the time of execution rather than being tightly integrated into the job definitions, which permit users to encapsulate the jobs on their own schedulers depending on the computational infrastructure. On our HPC resources, we achieve further flexibility by relying on the Balsam workflow toolkit \cite{salim2019balsam}. Using Balsam's Python programming interface, we populate a database with the desired pipeline actions and, using the specifics of the target computing resource, define a collection of jobs to be run. Balsam manages the execution of these jobs on the target computing resource, optimizing for concurrency and throughput, handling errors, and providing monitoring and reporting details as the pipeline jobs are executed. This level of control frees the user from laborious management of the compute jobs and enables Balsam to systematically manage execution and data management ensuring that the compute resources are used efficiently.

Pipeline users interact with Balsam via two interfaces: a Python API to define steps in the workflow and a command line interface for allocating resources and launching applications at the appropriate scale.  The mapping of parallel tasks to MPI ranks varies across applications: rank/section for montage with TraekEM2, rank/section-pair for alignment with AlignTK, and rank/subvolume for segmentation using FFN. This interface is mirrored in the Python/Balsam interface. Input data and configuration details are provided via Python calls, and passed to Balsam, which handles defining jobs in the underlying job database. 

Once the job database has been populated , one can use the Balsam command-line interface to submit jobs to the Theta queues for execution and to monitor the jobs as they run.

\section{Results}

In our experiments we used two HPC resources: Theta, an 11.69 PFLOPs supercomputer and Cooley, a GPU cluster with Nvidia K80s, both at the ALCF. Theta is composed of 4392 compute nodes, each with a 64-core, 1.3-GHz Intel Xeon Phi 7230 processor, 192GB DDR4 RAM and 16GB high-bandwidth MCDRAM. When needed, we also used a workstation with dual Xeon E5 2630v4, 256GB memory and two Titan X Pascal GPUs. Our usage of FFN relied on TensorFlow 1.14.

The microscope used on our experiments was the Zeiss SEM Gemini 300 \cite{carl2011detection}, which can provide images at up to 6nm of spatial resolution. The acquisition automation is done by the Atlas software from Zeiss.

The sample tissue was dissected from a 14 day old, post-natal mouse brain in primary visual cortex layer 4, and was prepared according to the protocol described by Hua \textit{et al.} \cite{hua2015large}. The tissue was then cut into 40 nm slices on ATUM\cite{schalek2011development} and scanned with the microscope at 6 nm resolution and 3.5 \textmu s pixel dwell time. For each of the 1312 slices, two 10833 x 14000 pixel tiles were scanned in sequence with 5\% overlap.

\subsection{Workflow Validation}
As an initial validation step, two stages of the pipeline--montage and alignment--were run on a subset of the data, using the Balsam execution backend. For montage, 128 sections of data were selected, with a corresponding Balsam job describing the input and configuration for TrakEM2. These jobs ran on 32 Theta nodes, with Balsam managing the distribution of work to compute nodes as they became available. We demonstrate this approach in the current case because this flexible approach to computing will become essential in the context of larger image stacks in the future. A similar approach was taken to run image alignment on this 128-image stack. For alignment, the database was populated with jobs to run on 16 nodes of the Cooley visualization cluster, with jobs distributed to compute nodes as they became available.

To highlight user interactivity with the pipeline, we provide a Jupyter notebook example where, given a raw dataset, the pipeline stages described above are submitted to a Balsam database with standard (or user-provided) configurations for execution on Theta. This approach permits the user to run collections of jobs multiple times, such as to perform parameter sweeps. We created  Balsam jobs for the TrakEM2 montage step with multiple configurations, varying the maximum and minimum octave used in the search for the correct overlap between two images, which also affects the run time, and calculated the error rate of those configurations on a given dataset (as shown on Table \ref{tab:trakemtimes}). An initial analysis of the resulting images was conducted to identify montage failures, using the image size as a proxy. The accumulated error was calculated by the number of images that were corrected by  changing the parameters. The final accumulated error shows the fraction of images that could not achieve a correct montage with any of the tested parameter sets; these, therefore, must be corrected through direct user intervention. We continue to develop metrics for identifying montage errors, in the interest of further automating this process.
\begin{table}[h!]
\centering
\resizebox{\columnwidth}{!}{%
\begin{tabular}{@{}lllll@{}}
\toprule
TrakEM2 - Min & TrakEM2 - Max & RunTime & Error Rate & Accumulated Error \\ \midrule
400           & 2000          & 100min      & 35\% & 35\%      \\
400           & 3000          & 260min      & 15\% & 10\%      \\
400           & 3500          & 450min      & 9\%  & 6\%      \\
1000          & 3500          & 520min      & 6\%  & 1\%      \\ \bottomrule
\end{tabular}}
\caption{Execution times of TrakEM2 headless montage macro on a test dataset (6x2 tiles of 15000x15000 pixels and 1128 slices divided into 8 different folders that represent the acquisition sessions). Each line shows the results of 8 Balsam jobs with 32 nodes, 4 ranks per node and the values for the TrakEM2 parameters for the minimum and maximum octaves used by the montage macro script. }
\label{tab:trakemtimes}
\end{table}

\begin{figure*}[tb!]
    \centering
    \begin{subfigure}[t]{0.6\textwidth}
        \centering
        \includegraphics[trim={186px 0 27px 0},clip,height=250px]{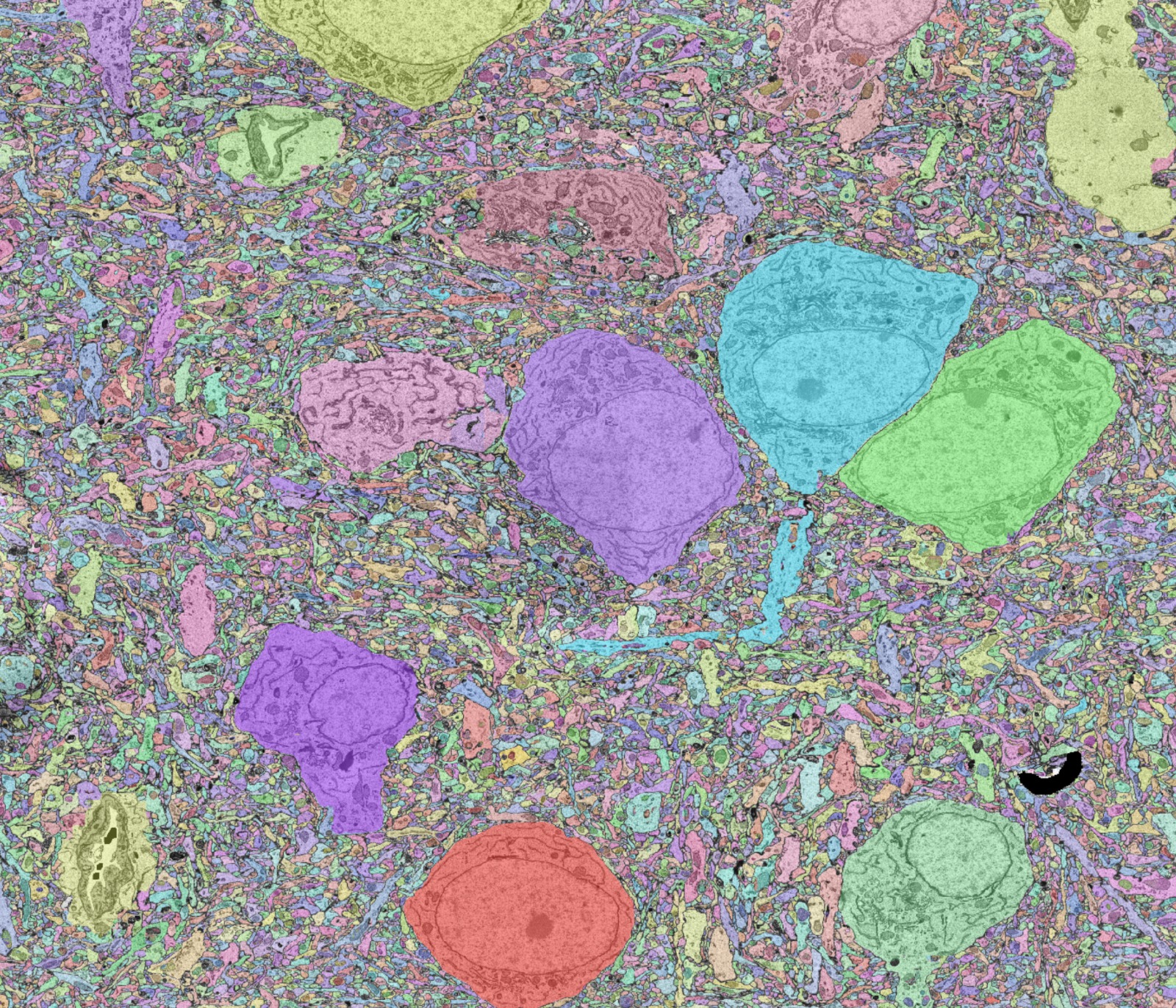} 
        \label{fig:ng2}
    \end{subfigure}
    \hfill
    \begin{subfigure}[t]{0.38\textwidth}
        \centering
        \includegraphics[height=250px]{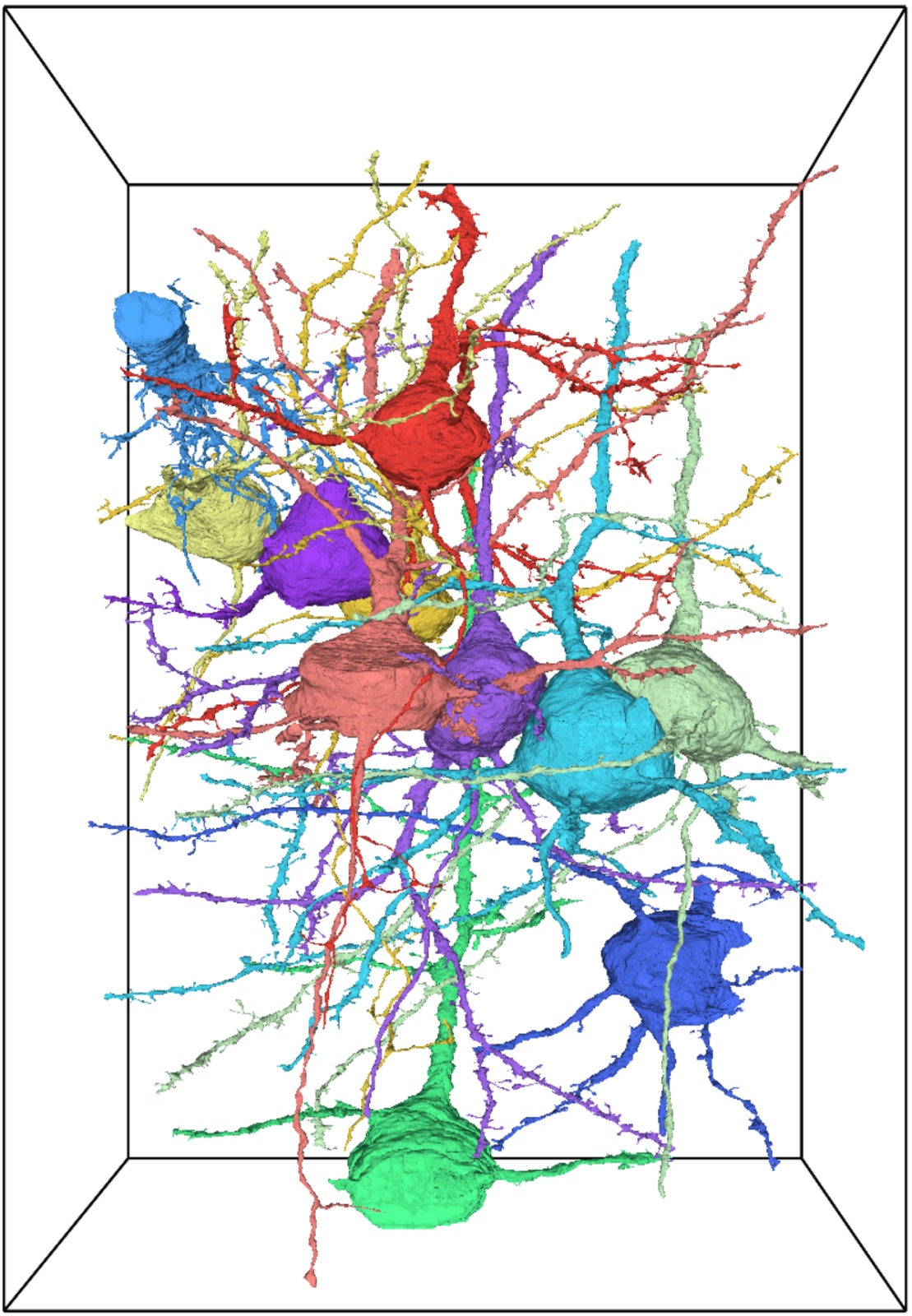} 
    \end{subfigure}
        \caption{Final visualization using Neuroglancer. On the left, an example of a raw image (black and white) overlapped with the inference labels(colors). On the right, 3D rendering of the same cells.}
\label{fig:neuroglancer_hanyu}
\end{figure*}

Lastly, we simulated online processing of images from the electron microscope by triggering a transfer of images on a schedule that approximates typical operation. For the dataset described in this work, each section was imaged as two separate tiles, each 8-bit tile having image dimensions 10833x14000 and occupying 151MB. The imaging time for each tile is on the order of 10 seconds, so a full section is imaged every 20 seconds. In this simulation, we transferred a full section from the microscope-connected machine to Theta every 20 seconds and added a montage job to the Balsam database, continuously, over a period of three hours. In the current paradigm, at this rate, a wafer of 200 sections would be imaged in about one hour, producing 30GB; this equates to a daily rate of 720GB. It is clear from this experiment that Theta is able to keep pace with the incoming jobs at this rate (each TrakEM2 job was run on a single Theta node, using 64 cores per node, and 2 threads per core, as we determined this to be the optimal configuration, with runtime averaging 440 seconds). To achieve this, we began with an initial allocation of Theta nodes, with the Balsam executor configured to grow and shrink the pool of nodes as needed, corresponding with the flow and ebb of incoming jobs. This demonstration is, in itself, not a compelling demonstration of the full extent of the current capability; it does, however, show that we have the technology in place to trigger image transfer and job injection when a section has been imaged, which will become a necessity in the future, where we anticipate transfer of images from multiple microscopes simultaneously to process on ALCF supercomputers. We are currently undertaking a scaling study to examine bounds on throughput in this scenario, which will be the focus of a future publication.

\subsection{Complete Pipeline}

We demonstrate the pipeline being executed from raw tiles to the final reconstruction on a 90 x 125 x 52 $\mu m$ volume of neural tissue. Each part of the pipeline was executed as a standalone call from a bash script using our Python wrappers or, where appropriate, by calling applications directly. 

After montage and alignment, the data size was 15000x20800x1312 voxels at 6x6x40 $nm^3$ resolution, in 8 bit grayscale, with a total size of 324 GB. Segmentation was carried out at 2x lower resolution to reduce merge errors and to increase speed. To perform training, we acquired an FFN model trained on the Kasthuri11 \cite{kasthuri2015saturated} dataset from the authors of \cite{januszewski2018high} as an initial checkpoint. Using manual annotations of a 256x256x128 voxel volume from our own dataset, we incrementally trained the base model until accuracy saturated at 0.91; this training was run on a separate workstation with dual Titan X GPUs for 12 hours, as a transfer learning job that didn't require Theta-scale computing. Before proceeding to inference, we first performed cell-body and vessel masking with U-Net/watershed on the workstation at 4x downsampled resolution, and used that as an initial segmentation. We then split a total volume of 6700x9900x1312 voxels (at 12x12x40 $nm^3$ resolution) into 3618 512x512x128 cubes with 32x32x16 overlap in each dimension. Inference jobs with the trained model were run on 32 nodes of the Cooley cluster each with 2 NVIDIA K80 GPUs, with one MPI rank per GPU, for a total of 72 hours; afterwards the subvolumes were reconciliated (recombined into a full volume) on a workstation for final visualization and error checking. Figure \ref{fig:neuroglancer_hanyu} shows the visualization of the reconstructed data in Neuroglancer. Our pipeline have been tested on datasets as large as 1Tb and it can be scaled to larger volumes as long as the resource allocation allows it. The limitations come from data sharing between different resources and the final user. Another limitation comes from the increase in the error rate of individual algorithms in larger volumes.

\section{Closing Remarks}
In summary, HAPPYNeurons provides a software pipeline to integrate electron microscopes with HPC facilities. We demonstrate an end-to-end connectomics reconstruction pipeline using HPC resources. This is achieved by wrapping multiple libraries as a coherent set of operations and providing the ability to chain them together using Balsam to enable more optimized scheduling on supercomputers. 
Due to the modular design of the workflow, the wrapped applications can be combined according to the needs of the current application and dataset, and new modules can be added with ease. HPC facilities can be used in a seamless manner, enabling the processing of large scale data without the monetary burdens of cloud computing. This integration paves the way for using supercomputers for connectomics reconstruction, in preparation for the deluge of data anticipated from faster next-generation microscopes, and to enable exascale computers to process it. This modular design also allows for different processes of the pipeline to target different accelerators on the HPC resources (i.e. dedicated GPU's). We are currently studying the execution of our pipeline at larger scale on Argonne supercomputers; these results will appear in a future publication. 

\subsection*{Acknowledgements}

We thank all members of the Kasthuri Laboratory at University of Chicago for providing the insights into Electron Microscopy and biology. We thank Dr. Shuichi Shigeno for data used in the preliminary tests.
We also thank Michal Januszewski for sharing a pre-trained FFN model and insights into FFN's hyper-parameters. This research is funded in part by, and used resources of, the Argonne Leadership Computing Facility which is a DOE Office of Science User Facility supported under Contract DE-AC02-06CH11357 and funded in part by NIH: 5R01MH110932-04.

\bibliographystyle{unsrt} 
\bibliography{references}

\end{document}